\documentclass[%
 amsmath,amssymb,
 aps,
]{revtex4-2}

\usepackage{newtxtext}
\usepackage{newtxmath}
\usepackage{amsmath}
\usepackage{array,booktabs,tabularx}
\usepackage{tabularx}
\usepackage{tabulary}
\usepackage{graphicx}
\usepackage{makecell}
\usepackage{hyperref}
\usepackage{geometry}
\usepackage{bm}
\usepackage{xcolor}

\newcolumntype{C}{>{\centering\arraybackslash}X}
\begin{document}

\preprint{APS/123-QED}

\title{Mixing Fronts in Smooth Chaotic Flows}

\author{Joris Heyman}
\email{joris.heyman@univ-rennes.fr}

\author{Tanguy Le Borgne}%
 \affiliation{Univ Rennes, CNRS, Geosciences Rennes, France.}

\author{Daniel Lester}%
\affiliation{School of Chemical and Environmental Engineering, RMIT University, 3000 Melbourne,
Victoria, Australia.  }

\date{\today}

\begin{abstract}
    Scalar mixing fronts develop at the interface of agitated fluids of different solute concentrations. In such fronts, scalar fluctuations form at both microscopic and macroscopic scales, due to stretching-enhanced molecular diffusion and hydrodynamic dispersion respectively.  
    While these two elementary processes are well understood separately, predicting how their coupling governs the evolution of concentration statistics within dispersing fronts remains a challenge.
	
    Here, we propose a theoretical framework to describe scalar fluctuations in fronts mixed by smooth chaotic flows.
    
    We find that the transfer of energy between the macroscopic and microscopic scalar fluctuation scales operates at a characteristic length scale $s_i$, for which dispersion and stretching-enhanced diffusion are of equal strength.
  
  	This leads to a closed expression for the concentration variance, which captures the results of direct numerical simulations with no fitting parameters, for a broad range of Péclet numbers.
    These findings open a new avenue for predicting both conservative and reactive mixing in smooth chaotic flows such as porous media or microfluidic flows. 
    
\end{abstract}

\maketitle

\section{Introduction}

\begin{figure}
\includegraphics[width=\linewidth]{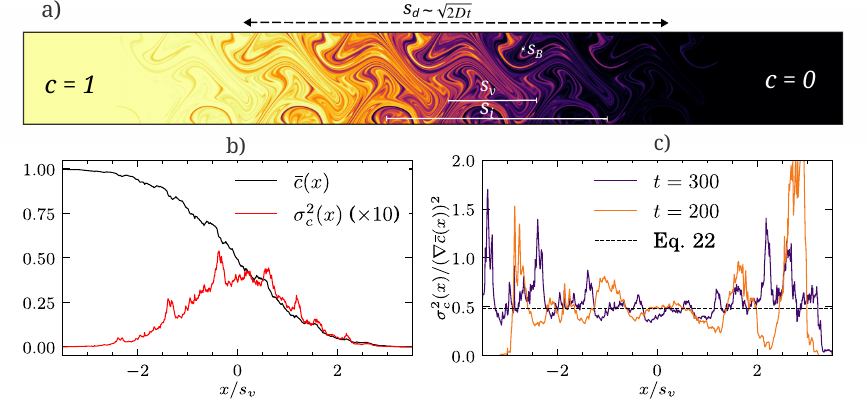} 
	\caption{a) Horizontal scalar front mixed and dispersed by a smooth, single-scale, and periodic chaotic flow in an infinite domain (sine flow). Scalar concentrations (colorscale) obey Eq.~\eqref{eq:ADE} with $\kappa=5\cdot10^{-6}$  and $\boldsymbol{u}$ defined by Eq.\eqref{eq:sineflow}, with $U=0.2$ ($\gamma\approx {\gamma^\prime}\approx  0.05$), and $t=300$.  The mixing scales appearing in mixing fronts are shown at real scale in the plot (from small to large). For the considered scenario, the Batchelor scale and its variant are $s_B\approx s_B'\approx 0.01 s_v$, the velocity scale is $s_v\equiv 1$, the injection scale is $s_i \approx 2.7 s_v$. b) Mean and variability of concentrations along the horizontal direction $x$. Averages are taken over a time window of 40, and along the vertical direction. c) Renormalization of the concentration variability by the mean gradient squared, together with the prediction of Eq.~\eqref{eq:variance}}
    \label{fig:front}
\end{figure}

Solute mixing by chaotic stirring of the fluid occurs in a range of natural and industrial processes, including static mixers~\citep{stroock2002chaotic},  turbulence ~\citep{villermaux2006coarse}, or porous media~\citep{HeymanPRL2020}. 
Away from the solute source, mixing fronts naturally develop at the interface between the diluting and dispersing solute and the surrounding fluid, forming scalar gradients over a wide range of length scales  (Fig.~\ref{fig:front}). Such interfaces are commonly observed in the fringe of contaminant plumes in laminar flows through porous media~\citep{rolle2019mixing}, or in turbulent density or thermal currents in lakes, coastal areas, or river junctions~\cite{horner2015mixing}.  Describing scalar fluctuations in mixing fronts is of particular importance in the context of mixing-driven reactions, because reactivity depends on the mixing state of chemical species~\citep{valocchi2019mixing,fox2003computational}.
While early-time lamella dilution~\citep{duplat2008mixing} and asymptotic mixing in confined domains~\citep{haynes2005controls,tsang2005exponential} are now well characterized, the multiscale evolution of mixing in expanding scalar fronts is not totally understood yet.

Here, we aim at predicting scalar fluctuations in mixing fronts agitated by \textit{smooth} chaotic flows, for which velocity fluctuations arise on a single, well-defined scale, denoted $s_v$. These laminar flows are common in porous materials at the pore scale~\citep{heyman2024mixing}, or in microfluidic devices.
At macroscale, well above $s_v$, the spatio-temporal evolution of mean scalar concentrations are well described by effective Fickian dispersion models~\citep{whitaker2013method}, at least in smooth chaotic flows of integrable velocity statistics. However, modeling the macroscale evolution of scalar concentration moments of higher order than the mean involve unclosed terms that depend on the microscale chaotic mixing process. While several empirical closure for these terms have been proposed in the past~\citep{fox2003computational}, a theoretical model based on the physics of the microscale mixing process is not available yet.
In turn, at microscale (below $s_v$) and close to a source of solute, scalar dilution in smooth chaotic flows is well predicted by Lagrangian lamellar models~\citep{balkovsky1999universal,Meunier2010,meunier2022diffuselet}. They capture the interplay of fluid deformation and molecular diffusion in the direction transverse to the solute filaments or sheets formed during fluid agitation. 
However, Lagrangian approaches do not allow for the prediction of asymptotic mixing regime when the elongated structures coalesce and merge, forming a continuously fluctuating scalar field. Aggregation rules~\citep{villermaux2003mixing,DuplatVillermaux08,heyman2024mixing} were used to describe the coalescence process and predict asymptotic scalar fluctuations decay in confined domains. 
However, such rules are difficult to apply to expanding fronts in ever-growing domains because aggregation becomes a function of both space and time~\citep{LeBorgneJFM2015}.

Microscale scalar mixing was also investigated with spectral models~\citep{Batchelor1959,kraichnan1974convection} through the description of the evolution of the power spectrum density of scalar fluctuations (PSD). When a macroscale source of scalar fluctuations exists, \citet{Batchelor1959} have shown that the scalar PSD exhibits, at low wavenumber $k$, a universal $k^{-1}$ decay caused by smooth fluid deformations, and at high $k$, an exponential cutoff caused by molecular diffusion. \citet{kraichnan1974convection} derived an analytical expression for the scalar PSD in the case of a smooth and random chaotic flow of infinitesimal time persistence. However, spectral approaches are limited to spatially smooth fluid deformations, and thus, are not directly applicable for mixing fronts developing scalar scales above $s_v$.

In particular, it is not fully understood how the spectrum of microscale fluctuations is mediated by dispersion in expanding fronts. \citet{tsang2005exponential} and  \citet{haynes2005controls} computed direct numerical simulation of scalar variance decay in two-dimensional time-varying sine-flows in periodic domains of variable size. They showed that, at low $k$, the shape of scalar fluctuations PSD depends on the domain size with respect to $s_v$, with a flat spectrum on small domains and a $k^{-1}$ spectrum in the limit of large domains. They also showed that the scalar variance decay is governed either by fluid stretching or by hydrodynamic dispersion in domains smaller or larger than $3 s_v$ respectively. The critical domain size separating the two regimes was obtained~\citep{haynes2005controls} by comparing the characteristic variance decay time caused by stretching enhanced diffusion to the characteristic variance decay time caused by a Fickian dispersive process. While the critical domain size distinguishes a \textit{local} and a \textit{global} control of scalar fluctuation decay in confined environments~\citep{haynes2005controls}, its role in the evolution of scalar front statistics in unconfined domains has not been considered so far. 

Here, we show that a similar length scale, denoted $s_i$, controls the magnitude of local scalar fluctuations arising in mixing fronts expanding in unconfined domains. In particular, we show that the incomplete mixing state of a dispersing scalar can be fully characterized by coupling, at the length-scale $s_i$, the macroscale dispersive transport laws~\citep{whitaker2013method} to microscale chaotic mixing spectral theories for smooth chaotic flows~\citep{Batchelor1959,kraichnan1974convection}. This leads to a theoretical prediction of the evolution of scalar variance within mixing fronts with no fitting parameters. Note that we do not consider turbulent flows, which generally differ from smooth chaotic flows.

The paper is organised as follows. In section 2, we first recall the conservation equations governing scalar variance evolution at macroscale and the theoretical shape of the scalar PSD at microscale. We then propose a theoretical closure for the scalar dissipation in mixing fronts based on the scale $s_i$. In section 3, we compare these analytical predictions with direct numerical simulations of scalar mixing fronts agitated by a smooth chaotic flow, the sine-flow~\citep{antonsen1996role,meunier2022diffuselet}. We show that the theory accurately captures the evolution of scalar variance in mixing fronts. Finally, we conclude on potential applications for modeling reactive transport processes. 

\section{Theory}

%
The transport of a scalar concentration field $c(\boldsymbol{x},t)$ under the action of an incompressible velocity field $\boldsymbol{u}(\boldsymbol{x},t)$ is governed by the advection diffusion equation
\begin{equation}
	\partial_t c + \boldsymbol{u} \cdot \nabla c = \kappa \nabla^2 c .
	\label{eq:ADE}
\end{equation}
where $\kappa$ is the diffusivity of the scalar and $\boldsymbol{u}$ is assumed here to originate from a smooth chaotic flow which stirs the scalar on a well defined length scale $s_v$. 

As shown in Fig~\ref{fig:front}a, this microscale agitation leads both to the dispersion and mixing of an initially sharp scalar front, expanding it to larger and larger distances and progressively reducing its sharpness. Because  mixing is not instantaneous, local microscale scalar fluctuations persist below $s_v$ (Fig~\ref{fig:front}b). Remarkably, the magnitude of these fluctuations is proportional to the square of the local mean concentration gradient, with a proportionality constant that appears independent of time and space (Fig~\ref{fig:front}c). 
The value of this constant has not been described yet in the case of dispersing front mixed by a smooth chaotic flow. The objective of this study is thus to provide a physical modeling of this proportionality constant, and investigate its dependency on the flow agitation properties and the diffusivity of the scalar. Beforehand, we recall macroscale and microscale theories of scalar spreading and mixing, then showing how they can be used to describe the overall mixing front scalar statistics.

\subsection{Background}
\subsubsection{Macroscale evolution of concentration mean and variance in a dispersing front}
Taking a Reynolds decomposition of the scalar concentration $c$ and velocity $\boldsymbol{u}$, we have 
\begin{eqnarray}
    c(\boldsymbol{x},t) &=& \bar{c}(\boldsymbol{x},t) + c'(\boldsymbol{x},t),
     \label{eq:reynolds_c} \\
     \boldsymbol{u}(\boldsymbol{x},t) &=& \bar{\boldsymbol{u}}(\boldsymbol{x},t) + \boldsymbol{u}'(\boldsymbol{x},t),
     \label{eq:reynolds_u}
\end{eqnarray}
where the averaging operator $\bar{\bullet}$ is the ensemble mean of the microscale chaotic agitation process, of length-scale $\leq s_v$. $\bar{\bullet}$ can also be interpreted as the result of a low-pass filtering with characteristic scale $s_v$. 
%
%
Inserting \eqref{eq:reynolds_c}-\eqref{eq:reynolds_u} in \eqref{eq:ADE}, and averaging gives 
\begin{equation}
	\frac{\partial \bar{c}}{\partial t} + \bar{\boldsymbol{u}}\nabla \bar{c} =  \nabla \cdot ((\kappa+\boldsymbol{D}) \nabla \bar{c}),
	\label{eq:meanconcentration}
\end{equation}
with $\bar{\boldsymbol{u}}$ the mean flow velocity over the microscale agitation and $\boldsymbol{D}$ is the dispersion coefficient obtained with the assumption of a Fickian macroscale flux expression~\citep{kapoor1997advection}
\begin{equation}
	\overline{\boldsymbol{u}'c'} \approx - \boldsymbol{D}\nabla \bar{c}.
\end{equation}
When microscale velocity fluctuations are isotropic, the dispersion tensor is also isotropic and expressed as $\boldsymbol{D}=D\mathbf {I}$. Owing to the Taylor-Green-Kubo relation between velocity fluctuations and dispersion, $D$ generally scales as 
 \begin{equation}
     D \sim U^2 T.
     \label{eq:defdisp}
 \end{equation}
 with $U$ and $T$ a characteristic flow agitation velocity and time.  
The dispersive limit is reached for any chaotic flow of finite correlation time and integrable statistics~\citep{whitaker2013method,dentz2023mixing}.  Dispersion leads to the spreading of solute and the formation of scalar gradients at increasingly large scales 
    $s_d \sim \sqrt{Dt}$
in mixing fronts (Fig.~\ref{fig:front}). Thus, we may assume an asymptotic separation of length and times scales between the mean field $\bar{c}(\mathbf{x})$ and the fluctuation $c'(\mathbf{x},t)$, justifying the use of the low-pass filter and Reynolds decomposition. 

The fluctuating part of the concentration field satisfies the inhomogeneous ADE
\begin{equation}
	\partial_t c' + \boldsymbol{u} \cdot \nabla c' - \kappa \nabla^2 c' = f(\mathbf{x}),\quad f(\mathbf{x})\equiv -\partial_t \bar{c} -\boldsymbol{u} \cdot\nabla \bar{c} - \kappa \nabla^2 \bar{c},
    \label{eq:ADE_fluc}
\end{equation}
As shown in \citet{kapoor1997advection}, multiplying Eq.~\ref{eq:ADE_fluc} by $c'$ and averaging, one may obtain an evolution equation for the variance of concentration fluctuations $\sigma_{c}^2=\overline{c' c'}$
\begin{equation}
	\frac{\partial \sigma_c^2}{\partial t} + \bar{\boldsymbol{u}}\nabla \sigma^2_c - (\kappa+D) \nabla^2 \sigma^2_c= -2 \kappa \overline{(\nabla c')^2}+ 2 D (\nabla \bar{c})^2,
    \label{eq:variance}
\end{equation}
where it is assumed that $\overline{\boldsymbol{u}c'^2}$ follows a Fickian macroscale flux ($- D\nabla\sigma^2_c$) and that $\overline{f}$ is negligible compared to the magnitude of microscale fluctuations. 
The left hand side of Eq.~\eqref{eq:variance} expresses the transport of scalar variance through advective, diffusive and dispersive fluxes while the right-hand side expresses the dissipation and production of scalar variance due to molecular diffusion and dispersive motion respectively.
The production of scalar variance is caused by large-scale dispersive fluxes arising from gradients in the mean concentration $\bar{c}$, which bring unmixed fluid patches in contact with each other. In contrast, the destruction of the scalar variance results from microscale diffusive fluxes. This term is not closed because it relies on sub-scale statistics of the scalar field. Here, we aim at providing a physically-based closure for microscale scalar dissipation in the case of a freely dispersing front mixed by a smooth chaotic flow.

\subsubsection{Microscale power spectra in a smooth chaotic flow}
As noted by \citet{Batchelor1959}, the stretching action of a smooth chaotic flow transfers scalar fluctuations from $s_v$ towards the Batchelor scale $s_B=\sqrt{\kappa/\gamma}$ with $\gamma$ is the mean stretching rate of the chaotic flow, defined as
\begin{equation}
	\gamma = \lim_{t\to \infty} \frac{1}{t} \overline{\ln \rho(t)}, 
\end{equation}
where $\rho(t)=\ell / \ell_0$ is the elongation of an infinitesimal fluid parcel of initial length $\ell_0$ by the flow.
The value of $\gamma$ is dictated by the accumulation of random fluid deformations produced by microscale velocity gradients, scaling as $U/s_v$ where $U$ is a characteristic velocity magnitude. $s_B$ is the length scale at which fluid compression is balanced by molecular diffusion, and thus the minimum scale at which scalar fluctuations can persist.
In the case of a uniform stretching, \citet{Batchelor1959} described the evolution of the scalar PSD $E_k(k)\equiv|\tilde{c}(k)|^2$, where the tilde symbol stands for the spatial Fourier transform
\begin{equation}
	\tilde{c} = \int c'(\boldsymbol{x})e^{i \boldsymbol{k} \cdot \boldsymbol{x}} \text{d}\boldsymbol{x}
\end{equation}
with $k\equiv |\boldsymbol{k}|=2\pi/s$ the wavenumber, inversely proportional to the length scale $s$. The theory rest on the observations that (i) the amplitude of scalar fluctuation is unchanged by the fluid deformation process, while their characteristic wavenumber increases exponentially because fluid stretching is compensated by transverse compression in incompressible flows and (ii) molecular diffusion acts as a low-pass filter at high wavenumbers.  \citet{kraichnan1974convection} generalized the spectral theory to smooth chaotic flows with random stretching of small persistence (main steps of the derivation are given in Supp. Mat. Section S1.). When a macroscale source produces scalar variance at a fixed rate $\chi_0$, \citet{kraichnan1974convection} shows that the microscale scalar PSD follows
\begin{equation}
	E_k(k,\chi_0) = \frac{\chi_0}{ \gamma k} {F}_{\gamma/{\gamma^\prime}}(2 k s_B') \exp(-2 k s_B'). 
	\label{eq:spectrum}
\end{equation}
where
$s_B'\equiv\sqrt{\kappa/{\gamma^\prime}}$ 
is a modified Batchelor scale defined with the fluctuating part of the stretching rather than the mean $\gamma$, expressed as
\begin{eqnarray}
	{\gamma^\prime} &=& \lim_{t\to \infty} \frac{1}{t} \overline{(\ln \rho(t)-\gamma t)^^2}.
	\label{eq:defgamma}
\end{eqnarray}
Note that both ${\gamma^\prime}$ and $\gamma$ have units of time frequencies. The function ${F}_{\gamma/{\gamma^\prime}}(k)$ is the solution of a second-order ODE with the ratio $\gamma/\gamma^\prime$ as parameter (see Supp. Mat. Section S1 for details and Eq.~(5.14) in \citet{kraichnan1974convection}). 

Eq.~\eqref{eq:spectrum} shows that the PSD decays as $k^{-1}$ for low wavenumbers, with an exponential cutoff at high wavenumbers. Note that the Kraichnan spectrum is comparable to Batchelor's prediction~\citep{Batchelor1959} obtained in the case of a quasi-static (long persistence) straining motion, with the difference that the latter has a steeper cutoff ($E_k \sim \exp(-k^2s_B^2)$) at high wavenumbers due to the absence of stretching fluctuations. Eq.~\eqref{eq:spectrum} is exact in the limit of (i) random flows of weak persistence with  (ii) spatially smooth stretching fields. These two conditions are discussed in the following. Flow persistence can be quantified by the Kubo number~\citep{sokolov2000ballistic,dentz2025coupled}, $Ku= {U T}/{s_v}$, where $T$ is the characteristic persistence time $T$ of velocity gradients. When $Ku\ll 1$, the mean stretching rate is expected to scale as
\begin{equation}
	\gamma \sim T\, (U/s_v)^2.
	\label{eq:stretching_def}
\end{equation}
Note that in weakly persistent flows, ${\gamma^\prime}$ is related to $\gamma$ through the dimension $d$ of the domain where mixing occurs~\citep{kraichnan1974convection,meunier2022diffuselet}:
\begin{equation}
	d / 2= \gamma / {\gamma^\prime}.
	\label{eq:dimension}
\end{equation}
In particular, when $d=3$, $\gamma=3{\gamma^\prime}/2$ and ${F}_{3/2}=1$. In turn, when $d=2$, $\gamma={\gamma^\prime}$, $s_B=s_B'$, and ${F}_{1} \approx 1+O(k^{1/2})$~\citep{kraichnan1974convection}.  
The assumption of the smoothness of the stretching field implies that $\gamma$ and ${\gamma^\prime}$ are spatially uniform and independent of scale. Spectral theories thus only captures scalar fluctuations at length scales smaller than $s_v$, where the velocity gradients can be considered to be approximately constant. Hence, they cannot describe the evolution of mixing fronts above $s_v$, since $\gamma$ drops to 0 and the flow agitation becomes essentially dispersive (Eq.~\eqref{eq:variance}). Owing to Eq.~\eqref{eq:meanconcentration}, the macroscale scalar PSD of mixing fronts is expected to follow a diffusive cutoff at large $k$
\begin{equation}
    E_{k,\bar{c}}(k) \sim \exp(- 2 (D+\kappa) k^2 t).
    \label{eq:spectrum_dispersive}
\end{equation}
In the following, we show that the description of mixing at macroscale (Eq.~\eqref{eq:variance}) and at microscale (Eq.~\eqref{eq:spectrum}) can be matched at the critical length scale $s_i$~\citep{tsang2005exponential,haynes2005controls}, called \textit{injection} scale, at which the macroscale scalar fluctuations are transmitted to higher wavenumbers at a rate $\chi_0$ controlled by dispersion.

\subsection{Scalar variance in mixing fronts}
Since $s_d$ grows in time and $s_v$ and $s_B$ are fixed, mixing fronts have the characteristic asymptotic separation of lengthscales
\begin{equation}
	s_d \gg s_v \gg s_B,
\end{equation}
Thus, the macroscale dispersion and microscale mixing theories presented above are valid in the limits of large and small scales respectively. 
Via the Plancherel formula, the microscale scalar variance in a freely dispersing front can be obtained by integrating Eq.~\eqref{eq:spectrum}. To avoid divergence at low wavenumbers due to the $k^{-1}$ scaling, one has to provide a minimum integration bound such that 
\begin{eqnarray}
	\sigma^2_{c} \approx \frac{1}{\pi} \int_{2\pi/s_i}^{\infty} E_k(k,\chi_0) \, \text{d} k.
	\label{eq:integral}
\end{eqnarray}
The minimum bound $k=2\pi/s_i$ is the wavenumber at which macroscale scalar fluctuations are \textit{injected} into the microscale agitation flow. We thus call this specific length scale the \textit{injection} scale. The physical meaning and values of the scalar variance injection rate $\chi_0$ and the injection scale $s_i$ in unbounded mixing fronts are discussed in the following.

\subsubsection{Scalar variance injection rate $\chi_0$}
Since scalar fluctuations are produced at large scales by the mean dispersive flux (Eq.\eqref{eq:variance}), the variance injection rate $\chi_0$ must also be controlled by the macroscale dispersion process. Note that this hypothesis is equivalent to the \textit{spectral equilibrium} hypothesis in turbulence, which assumes that the scalar dissipation rate is set by the inertial-convective range, regardless of the extent of the viscous-diffusive range~\citep{fox2003computational}. The variance production rate due to mean scalar gradients is
   $ {\text{d}\sigma^2_c}/{\text{d}t} = 2 D (\nabla \bar{c})^2$.
Assuming that this is entirely transmitted to smaller scales, then
\begin{equation}
     \chi_0 \equiv  \pi \frac{\text{d}}{\text{d}t} \sigma^2_c = 2 \pi D (\nabla \bar{c})^2.
     \label{eq:chi0}
\end{equation}
%

\subsubsection{Injection scale $s_i$}
As we have seen, at the microscale, the scalar fluctuations inherited from the macroscale obey a well-defined spectral structure (Eq.~\eqref{eq:spectrum}). The transfer of energy from macro- to micro-scales must thus operate at a well-defined length-scale $s_i$, which separates dispersive processes (well above $s_v$) from fluid deformation processes (well below $s_v$). 
Drawing on previous observations in confined domains~\citep{haynes2005controls,tsang2005exponential}, we define $s_i$ as the scale at which the characteristic time to destroy variance by dispersion $t_D$ is comparable to the characteristic time to destroy variance by stretching-enhanced mixing $t_\gamma$.
The characteristic time for variance decay by dispersion in a domain of length $s_i$ is 
\begin{equation}
	t_D^{-1}=2 \left(\frac{2\pi}{s_i}\right)^2 D
\end{equation} 
while the characteristic time for variance decay by stretching-enhanced mixing in the same domain is~\citep{Kalda2000}
\begin{equation}
	t_\gamma^{-1}=\frac{\gamma^2}{2 {\gamma^\prime}}.
    \label{eq:tgamma}
\end{equation} 
The first equation is obtained from the dispersive spectra (Eq.~\eqref{eq:spectrum_dispersive}),  while the second results from log-normal distribution of elongations, and can be found in a number of references~\citep{Meunier2010,thiffeaultbook,HeymanPRL2020}. Since $\gamma' = \gamma$ at small $Ku$ (Eq.~\eqref{eq:dimension}), Eq.~\eqref{eq:tgamma} simplifies to $t_\gamma^{-1}= d \gamma / 4$ in $d-$dimensional flow.
%
Equating $t_D$ and $t_\gamma$ sets the value of the injection length scale to
\begin{equation}
	s_i= 4\pi \sqrt{{D {\gamma^\prime}}/{\gamma^2}}.
	\label{eq:si}
\end{equation}
For weakly persistent flows in $d=2$ dimensions, we expect $s_i= 4\pi \sqrt{D/\gamma}$ and in  $d=3$ dimensions,  $s_i= 4\pi \sqrt{2D/(3\gamma)}$. Note the similarity between the expression of $s_i$ and $s_B$, but with the molecular diffusivity $\kappa$ replaced by the macroscopic dispersion coefficient $D$.  Note also that the magnitude of $s_i$ is different from, although proportional to, the velocity scale $s_v$. Indeed, using the expected scaling for $\gamma$ and $D$ with $U$ and $T$ (Eqs.~\eqref{eq:stretching_def} and \eqref{eq:defdisp} respectively) in Eq.~\eqref{eq:si}, we get $s_i \sim s_v$. The precise value of $s_i$ depends on the capability of a given heterogeneous flow field to disperse and to mix at the same time. 

As mentioned previously, the competition between dispersion and stretching-enhanced mixing has been already shown to be critical for the decay of scalar variance in confined domains~\citep{haynes2005controls,tsang2005exponential}. A similar competition was also mentioned by \citet{baldyga1992interactions} in turbulent flows, where the value of the ratio of dispersive time versus mixing time was found to determine the limiting process for mixing. However, to the best of our knowledge, the importance of $s_i$ in the context of unbounded mixing fronts was not known a-priori, so as its explicit expression Eq.~\eqref{eq:si}.

\subsubsection{Scalar variance}
The integral~\eqref{eq:integral} is analytical in 3D domains when ${F}_{\gamma/\gamma^\prime} = 1$. It yields
\begin{eqnarray}
     \sigma^2_c \approx \frac{2 D (\nabla \bar{c})^2}{ \gamma} |Ei( - 4\pi s_B'/s_i)|,
     \label{eq:variance_closure}
\end{eqnarray}
where  $Ei$ is the exponential integral.  In 2D domains,  the integral remains well approximated by Eq.~\eqref{eq:variance_closure} even if ${F}_{\gamma/\gamma^\prime}=1+O(k^{1/2})$ (see Supp. Mat. Fig.~S3). This is due to the importance of the low wavelength $k^{-1}$ scaling and the large wavelength exponential cutoff in $E_k$ rather than the weak power-law increase of $F$. We thus consider Eq.~\eqref{eq:variance_closure} to be approximately valid for all space dimensions.

It is important to note that when integrating Eq.~\eqref{eq:integral}, we have implicitly assumed that the scalar spectrum is well approximated by the Kraichnan model up to the scale $s_i$, given by Eq.~\eqref{eq:si}. This hypothesis is justified because $s_i$ is always of the order of $s_v$. While the scalar PSD likely deviates from the $k^{-1}$ scaling close to $s_i$, the comparison of Eq.~\eqref{eq:variance_closure} with numerical simulations presented in the following indicates that this has a limited impact on the accuracy of the prediction. 

Equations \eqref{eq:si} and \eqref{eq:variance_closure} provide a complete prediction for the magnitude of scalar variance within mixing fronts, based on four independent physical parameters: the molecular diffusion $\kappa$, the dispersivity $D$, the mean and variability of the stretching process, $\gamma$ and ${\gamma^\prime}$ respectively. Note the absence of any numerical constant that cannot be estimated independently. Eq.~\eqref{eq:variance_closure} can also be used to close the macroscale evolution equation of scalar variance~\eqref{eq:variance} as follows.  Assuming negligible variance transport in Eq.~\eqref{eq:variance}, we have an equilibrium between variance dissipation and production, e.g. $2\kappa \overline{(\nabla c')^2} \approx {2D \left( \nabla \bar{c} \right)^2}$. Moreover, upon dimensional arguments, the dissipation term can be written as
\begin{equation}
	2\kappa \overline{(\nabla c')^2} \sim \sigma^2_c / \tau_m,
    \label{eq:dissip}
\end{equation}
with $\tau_m$ a characteristic decay time of scalar variance. Replacing $\sigma^2_c$ by its estimate from microscale mixing theory (Eq.~\eqref{eq:variance_closure}), gives
\begin{equation}
	\tau_m = \gamma^{-1} \left| Ei(-4\pi s'_B/s_i) \right|.
	\label{eq:taum}
\end{equation}
Eqs.~\eqref{eq:taum}, \eqref{eq:dissip} and \eqref{eq:variance} can be used to predict the spatio-temporal evolution of scalar variance in mixing fronts stirred by smooth chaotic flows.

Defining a dimensionless Péclet number as
\begin{equation}
	Pe \equiv \frac{\gamma s_v^2}{\kappa} = \frac{s_v^2}{s_B^2},
\end{equation}
$\tau_m$ scales as $Ei(1/Pe^{1/2})$, a function that follows the logarithmic growth $\log(Pe)/2$ at asymptotically large Péclet number. Thus Eq.~\eqref{eq:taum} takes a form similar as the expression of lamellar mixing time in chaotic flows~\citep{Villermaux2019}, $\tau_m \sim \gamma^{-1} \ln Pe_0$, with $Pe_0=s_0^2\gamma/\kappa$ a Péclet number defined on the initial size of lamella $s_0$. In contrast to the lamellar mixing time, the characteristic time obtained in fully developed mixing fronts is independent of the initial thickness of scalar interfaces.

In the following section, we numerically test the accuracy of these theoretical predictions in a periodic, two-dimensional, smooth, and chaotic flow--the random sine flow--for which $D$, $\gamma$ and ${\gamma^\prime}$ can be estimated independently.

\section{Comparison to direct numerical simulations}
To test the previous theoretical developments, we consider scalar mixing fronts in a two-dimensional single scale flow field, the random sine flow, as a prototype of chaotic flow for which the persistence, the dispersivity and the stretching rate can be varied. Sine flows have been extensively used to investigate the key feature of chaotic mixing~\citep{Pierrehumbert:1994aa,antonsen1996role,meunier2022diffuselet}. Their relative simplicity makes them computationally tractable even for large domains and times.  Scalar transport is solved in two complementary scenarios (Fig.~\ref{fig:domain}): (i) a fixed mean scalar gradient in a small periodic domain and (ii) a freely dispersing front in an large aperiodic domain. 

\subsection{Velocity field}
The sine flow is a continuous transformation operating on a periodic domain $[0,s_v]\times [0,s_v]$, which is defined as $\boldsymbol{u}=(u,v)$, with
\begin{eqnarray*}
	v(x,t) &=& \left\lbrace \begin{array}{ll}
		U \sin( 2\pi x/ s_v + \phi_j)  &\text{ for  } j T< t< (j+1/2) T, \\
		0 & \text{ for  } (j+1/2) T< t< (j+1) T \\
	\end{array} \right. ,\\
	u(y,t) &=&  \left\lbrace \begin{array}{ll}
		0  &\text{ for  } j T< t< (j+1/2) T, \\
		U \sin( 2\pi y/ s_v + \psi_j) & \text{ for  } (j+1/2) T< t< (j+1) T \\
	\end{array} \right. ,\\
	\label{eq:sineflow}
\end{eqnarray*}
with $T$ the time period of the flow, $j$ the time period number and $U$ a constant velocity magnitude. $\phi_i,\psi_i \in [0,\pi]$ are random phases that change at each time period. 
The velocity field is divergence free. 
When $Ku = UT/s_v \to 0$ , the flow changes rapidly over time, causing minimal fluid deformations, while for $Ku \to \infty$ the persistence of the flow strongly affects the deformation of the fluid elements, which tend to align in the main shear directions.
The limit $Ku \to0$ is the $\delta$-time correlated flow envisioned in \citet{kraichnan1974convection}, where Eq.~\eqref{eq:spectrum} is valid. Without loss of generality, we set $T=1$ and $s_v=1$ such that the unique free parameter of the flow is $Ku \equiv U$, which controls both the stretching and dispersion statistics of the flow and its persistence. 
Particle dispersion in random sine flow is given in the long time limit by~\citep{tsang2005exponential}
\begin{equation}
	D\equiv \langle \Delta x^2+\Delta y^2\rangle/(2T) =   {U^2T}/{16}
	\label{eq:dispersion}
\end{equation}
where $\Delta x$ and $\Delta y$ are the mean squared displacement caused by a single flow period (see Supp. Mat. Section S2.A).
In the Kraichnan limit, the mean stretching rate is given by~\citep{meunier2022diffuselet,kraichnan1974convection} 
\begin{equation}
	\gamma= {\pi^2 U^2 T}/({8 \, s_v^2}),
	\label{eq:mean_stretching}
\end{equation}
while the variance of stretching rates is equal to the mean since the flow is two-dimensional, e.g. $\gamma'=\gamma$ (Eq.~\eqref{eq:dimension} with $d=2$). 
%

\subsection{Scalar transport}
We consider two mixing front scenarios: (i) a forced scenario where the mean macroscopic scalar gradient remains constant, and (ii) an unforced scenario where an initially sharp front progressively expands and flattens due to fluid agitation. The forced scenario is used as a computationally efficient tool to validate the theory in various flow and diffusion conditions, since it freezes the macroscopic front evolution to a steady state. In turn, the unforced scenario demonstrates that the theory is generally valid for dynamically evolving mixing fronts, which develop scalar gradients at scales much larger than $s_v$. Note that these two cases are different than previous studies~\citep{tsang2005exponential,haynes2005controls}, that considered scalar decay in confined periodic domains without persisting macroscale scalar gradients.

\begin{figure}
	\centering \includegraphics[width=\linewidth]{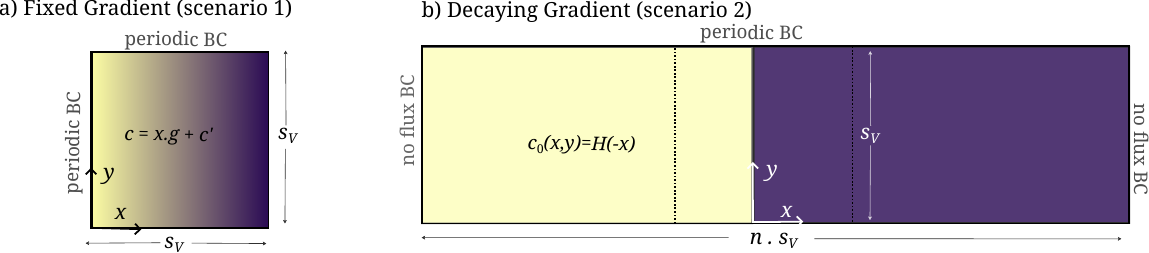} 
	\caption{Sketch of the two fronts studied numerically. a) Scenario 1 solving for $c'$ (Eq.~\eqref{eq:ADE_fluc}) with a mean macroscale scalar gradient $g$ enforced in direction $x$. Domain size is $[0,s_v] \times [0,s_v]$ and boundary conditions are periodic on both $x$ anad $y$. b) Scenario 2 solving for $c$ (Eq.~\eqref{eq:ADE}) with a sharp initial front in the $x$ direction which evolve in absence of any forcing in a long domain of size $[0,n s_v] \times [0, s_v]$, with $n$ a large integer number taken to be 10 in Figure~\ref{fig:macrodisp}. Boundary conditions are periodic in the $y$ direction and no-flux in the $x$ direction.}
	\label{fig:domain}
\end{figure}

The numerical domains and boundary conditions are described in Fig.~\ref{fig:domain} and in the following.
For the first scenario, the presence of a fixed large-scale gradient $\nabla \bar{c} \equiv \boldsymbol g$ is obtained by splitting the scalar field into mean and fluctuations
\begin{equation}
    c(x,t)= \bar{c}+c'=\boldsymbol{g}\cdot \boldsymbol{x} + c'(x,t)
    \label{eq:fluc}
\end{equation}
Inserting \eqref{eq:fluc} into the advection diffusion equation~\eqref{eq:ADE} leads to 
\begin{equation}
	\partial_t c' + \boldsymbol{u} \cdot \nabla c' = \kappa \nabla^2 c' -\boldsymbol{u} \cdot \boldsymbol{g}
    \label{eq:ADE_fluc_num}
\end{equation}
The presence of a source term in the advection-diffusion equation creates a forcing for scalar fluctuations on the scale of the velocity $\boldsymbol{u}$. Without loss of generality, we focus on a mixing front orientated in the $x$ direction, for which the forced gradient is 
\begin{equation}
	\boldsymbol{g} = g \, \boldsymbol{e}_x,
\end{equation}
thus leading to a source term equal to $- g u(y) \boldsymbol{e}_x$. 

In the forced scenario, we take periodic boundary conditions for $c^\prime$ on the unit square (Fig.~\ref{fig:domain}). Indeed, since the velocity field is periodic with period $s_v=1$, the scalar fluctuation field forced by a mean gradient is also periodic, and can thus be solved on the unit square. 
In the second scenario, since the mixing front spreads by dispersion, we must solve Eq.~\eqref{eq:ADE} on a long rectangular domain of size $[0,n]\times [0,1]$, where we choose $n=20$ (Fig.~\ref{fig:front} and ~\ref{fig:domain}b). The initial condition is taken as a sharp front 
\begin{equation}
	c_0(x,y) = \text{Heaviside}(-x).
\end{equation}
Periodic boundary conditions are set in axis $y$, while no-flux conditions are imposed in axis $x$.  

\subsection{Numerical scheme and convergence}
In both scenarios, the advection-diffusion equation is solved using spectral decomposition in Fourier space. 
Following the numerical method developed in \citet{meunier2022diffuselet}, unidirectional advection with source is solved exactly with Fourier transform in a single coordinate, while diffusion is solved in the full spectral domain. Denoting $\tilde{c}_x$ (respectively $\tilde{c}_y$) the Fourier transform of $c$ in direction $x$ (respectively $y$), and $\Delta t\leq T/2$ the numerical time step, the advection part in the first half-time period ($iT<t<(i+1/2)T$),
\begin{eqnarray}
    \frac{\partial \tilde{c}_x}{\partial t} + i k_x u(y) \tilde{c}_x = \mathcal{F}_x(- g\, u(y)) = - g \, \delta(k_x) u(y).
\end{eqnarray}
This yields
\begin{eqnarray}
    \tilde{c}_x (k_x,y,t+\Delta t)=  \tilde{c}_x(k_x,y,t) \exp(-ik_x u(y) \Delta t)-g \delta(k_x) u(y)\Delta t.
\end{eqnarray}
Diffusion is solved sequentially in the full frequency domain,
\begin{eqnarray}
    \tilde{c} (k_x,k_y,t+dt)=  \tilde{c}_x(k_x,k_y,t) \exp(- \kappa k^2 \Delta t).
    \label{eq:diffusion}
\end{eqnarray}
The second half period  ($(i+1/2)T<t<(i+1)T$) is solved similarly, this time without source term, with
\begin{eqnarray}
    \tilde{c}_y (x,k_y,t+dt)=  \tilde{c}_y(x,k_y,t) \exp(-ik_y v(x)\Delta t)
\end{eqnarray}
followed by a diffusive step~\eqref{eq:diffusion}.
The only approximation made by the numerical scheme is the operator splitting. Thus, the algorithm shows only first-order convergence in $\Delta t$, but with a mean relative error smaller than 0.5\% when using a large time step $\Delta t= T/2$. Herein, we set here $\Delta t= T/10$ to be below 0.1\% error. The domain is discretized by a regular grid with a grid size $\Delta x = s_B / 5$ fine-enough to resolve the expected minimum scale of scalar fluctuations. 

\begin{figure}[h]
    \centering
    \includegraphics[width=\linewidth]{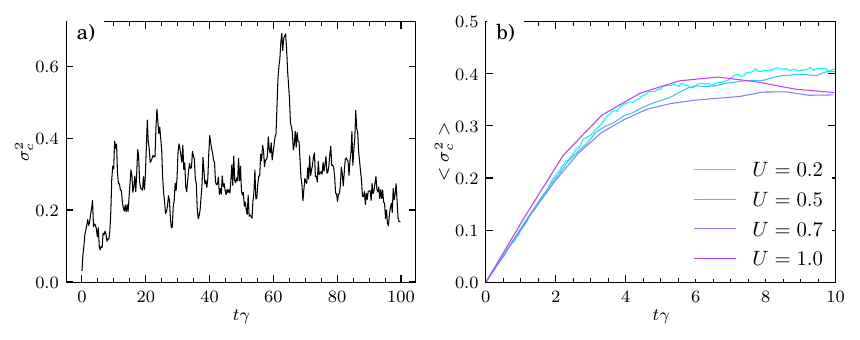}
    \caption{a) Time evolution of the scalar variance in the forced scenario for a single flow realization b) Corresponding time convergence of the mean $<\sigma^2_c>$ over 100 realizations of the random phases for various $U$ at fixed $s_B'=5.3 \cdot 10^{-3}$.}  
    \label{fig:time}
\end{figure}

In the forced scenario, we run simulations for a time large enough to reach stationarity (in a statistical sense) after which statistics are sample over long time periods.  Figure~\ref{fig:time}b shows that stationarity is achieved for $t \gamma \gtrapprox  10$, time after which the scalar variance oscillates around a mean constant value (Figure~\ref{fig:time}a). This is consistent with the time it takes for scalar fluctuations to be transferred from large to small scales, at the mean rate $\gamma$.  Statistical averages are then taken over time period larger than $40\gamma$.

\begin{figure}[h]
	\includegraphics[width=0.8\linewidth]{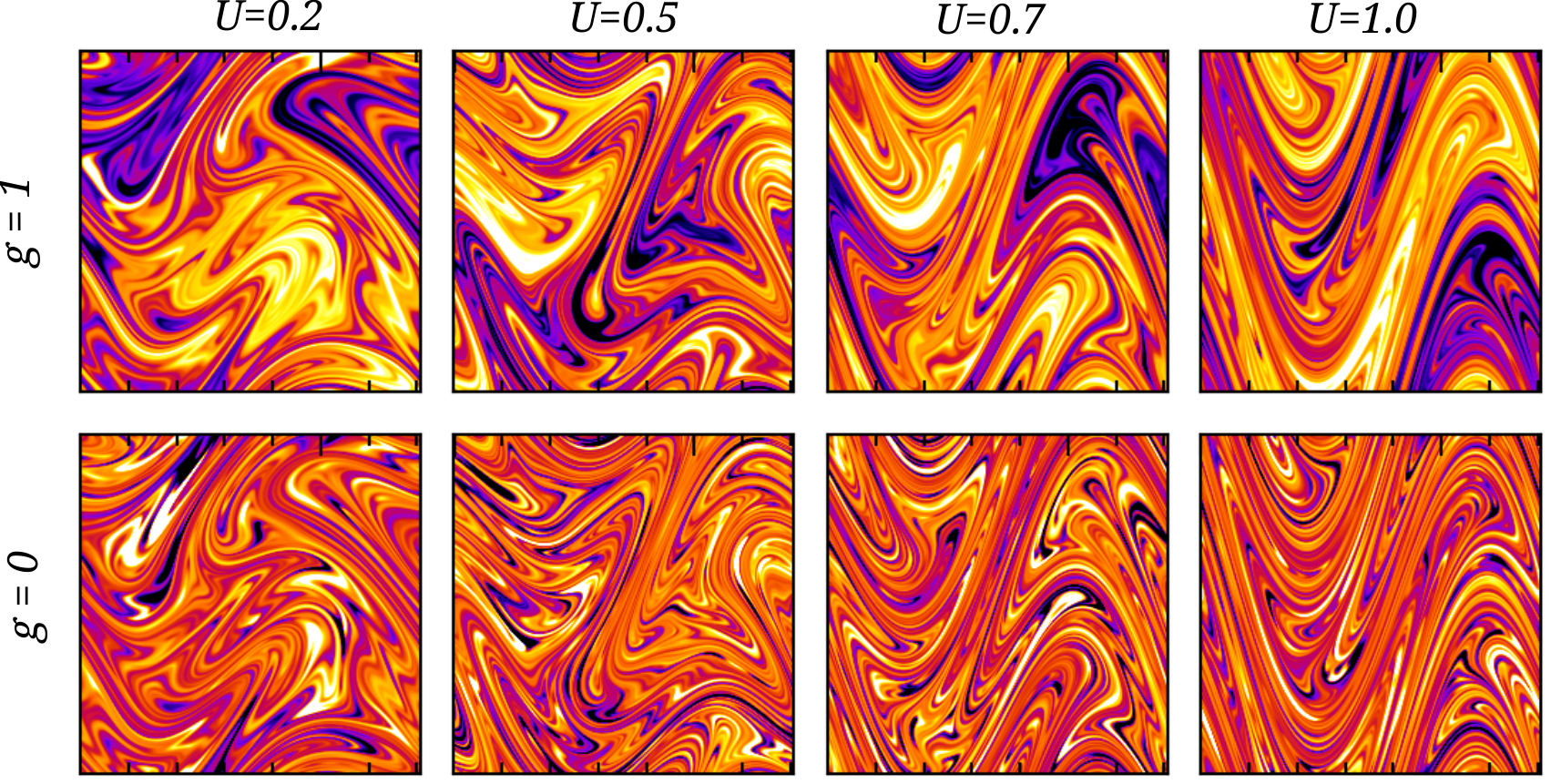}
	\caption{Concentration fields $c'$ for various amplitude $U$ for forced mixing $g\equiv \nabla\bar{c} =1$ (top) and unforced scalar decay $g=0$ (bottom) in a unit, periodic domain. In the scalar decay case, concentrations are rescaled by the standard deviation. We used a fixed ratio $\gamma/\kappa=5\cdot10^5$ in all simulations.}\label{fig:open_closed}
\end{figure}

\subsection{Scenario 1 : Fluctuations over a mean scalar gradient}\label{sec:fluct}
In the forced scenario, the simulated scalar field reaches a statistically steady state, previously reported as strange eigenmode~\citep{gouillart2009open} or persistent pattern~\citep{rothstein1999persistent}. 
As shown in Fig.~\ref{fig:open_closed}, the spatial structure of this asymptotic mixing state differs in the presence or absence of large-scale forcing ($g=0$ or $g=1$ respectively). When $g>0$, scalar fluctuations appears on a large range of scales, while for $g=0$, scalar gradients are confined at small scales. In turn, the persistence of the flow (increasing $U$) does not qualitatively affect the heterogeneity of the concentration field. 
In the following, we quantify this equilibrium state by investigating the scalar fluctuation spectrum, variance, and probability density function.  Statistical averages are estimated over time periods of $40 \gamma$.

\begin{figure}[h!]
	\includegraphics[width=\linewidth]{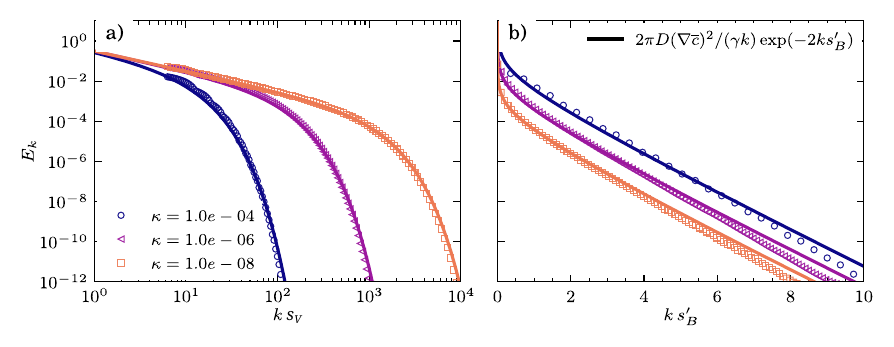}
	\caption{ 
		Comparison between power density spectrum obtained via numerical simulation (dots) and model (lines) combining Kraichnan spectrum (Eq.~\eqref{eq:spectrum}) with the injection rate set by dispersion (Eq.~\eqref{eq:chi0}). Flow amplitude is fixed to $U=0.1$ while molecular diffusivity $\kappa$ is varied. Spectrum is plotted in log-log with $k$ normalized by $1/s_v$ (a) and in log-lin scale with $k$ normalized by $1/s_B'$ (b).
	}
	\label{fig:spectrum}
\end{figure}

As shown in Fig.\ref{fig:spectrum}a, the power spectrum of microscale scalar fluctuations is well predicted by Kraichnan model~(Eq.~\eqref{eq:spectrum}), if the scalar variance injection rate is chosen to be the macroscale dispersive flux (Eq.~\eqref{eq:chi0}):
\begin{equation}
	\chi_0 = 2 \pi D (\nabla \bar{c})^2 = \pi U^2 T (\nabla \bar{c})^2 / 8
\end{equation}
where $D$ was obtained from \eqref{eq:dispersion}. Combining the expression $\chi_0$ with the expression of $\gamma$ (Eq.~\eqref{eq:mean_stretching}), yields a constant ratio
\begin{equation}
	\frac{\chi_0}{\gamma} = \pi^{-1} (s_v \nabla \bar{c})^2,
	\label{eq:chi0_sine}
\end{equation}
independent of the flow amplitude. Thus, in the limit of weakly persistent flows, $E_k\sim (\pi k)^{-1} (s_v \nabla \bar{c})^2$  for small $k$, independent of $U$ and thus of the persistence of the chaotic flow. 

Notably, we show in Fig.\ref{fig:spectrum}b that the characteristic exponential decay of the scalar PSD at large wavenumbers is well fitted by Eq~\eqref{eq:spectrum} in the case of weak flow persistence ($U=0.1$). In Supp. Mat. Section S2.C, we show that for higher flow persistence ($U>0.5$), the numerical spectrum exhibits a steeper cutoff at high frequencies compared to the exponential decay predicted by the Kraichnan model. Nevertheless, we will see that the scalar variance remains relatively well predicted at moderate flow persistence, since the majority of the fluctuation energy is concentrated at low frequencies, which are independent of flow persistence (Eq.~\eqref{eq:chi0_sine}).

\begin{figure}
 	\centering \includegraphics[width=\linewidth]{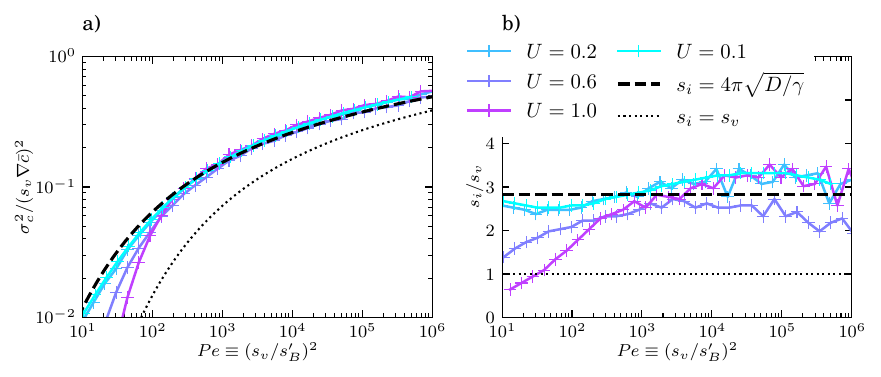} 
 	\caption{a) Dependence of the variance of scalar fluctuations $\sigma^2_c$ upon the Péclet number and flow amplitudes ($U$) in the presence of a mean scalar gradient (colored lines with markers). Numerical estimates from scenario 1 are plotted in colored lines with markers. The theoretical prediction (Eq.~\eqref{eq:variance_closure}) is plotted in dashed and dotted lines, with $s_i=4\pi\sqrt{D/\gamma} = 2\sqrt{2} s_v$ and $s_i=s_v$ respectively. 
 		 b) Numerical estimate of the lower integration bound $k=2 \pi/s_i$, for Eq.~\eqref{eq:integral} to match the numerical scalar variance.}
\label{fig:variance}
 \end{figure}

Having studied the scalar PSD, we now describe its integral measure, the scalar variance. In Fig.~\ref{fig:variance}, we compare the scalar variance at various Péclet number to the theoretical model proposed (Eq.~\eqref{eq:variance_closure}), with $s_i$ given by Eq.~\eqref{eq:si}. For the sine flow, this gives
\begin{equation}
	s_i \equiv 4\pi\sqrt{D / \gamma} = 2\sqrt{2} \, s_v \approx 2.8 \, s_v,
	\label{eq:si_sine}
\end{equation}
for weak persistence ($\gamma\approx\gamma')$. As shown on Figure.~\ref{fig:variance}a, Inserting this value in Eq.~\eqref{eq:variance_closure} accurately captures the numerical data over a large range of molecular diffusivity values and flow amplitudes.
For comparison, we also show the theoretical prediction for the trivial choice of the injection scale, $s_i=s_v$, which underpredicts the fluctuations by a factor 2 at $Pe=10^3$. In Fig.~\ref{fig:variance}b, we plot the exact numerical estimate of $s_i/s_v$ for the integral \eqref{eq:integral} to match the observed scalar variance, showing that the injection scale is relatively invariant across a large range of $Pe$, and aligns with the theoretical prediction of $s_i \approx 2.8 s_v$. 

The accuracy of Eq.~\eqref{eq:variance_closure} to predict scalar fluctuations in 2D mixing fronts is not straightforward, because it relies on the approximation that ${F}_{\gamma/{\gamma^\prime}} \approx 1$, which is strictly true in 3D. The accuracy of this approximate model is due to importance of the universal range $k^{-1}$ and the exponential cut-off in the integral of $E_k$, rather than the exact power-law scaling at intermediate wavenumbers (see also Supp. Mat. Fig.~S3). 
The divergence between the model and data at low $Pe$ for moderate flow persistence can be explained by the increasing role of the high-wavenumber cutoff in the integral, which departs from the exponential for persistent flows. In Supp. Mat. Fig.~S3, we show that taking into account the modification of the high-wavenumber cutoff in the spectrum with flow persistence allows to capture part of the variance drop observed below $Pe=200$.

\begin{figure}
	\centering \includegraphics[width=\linewidth]{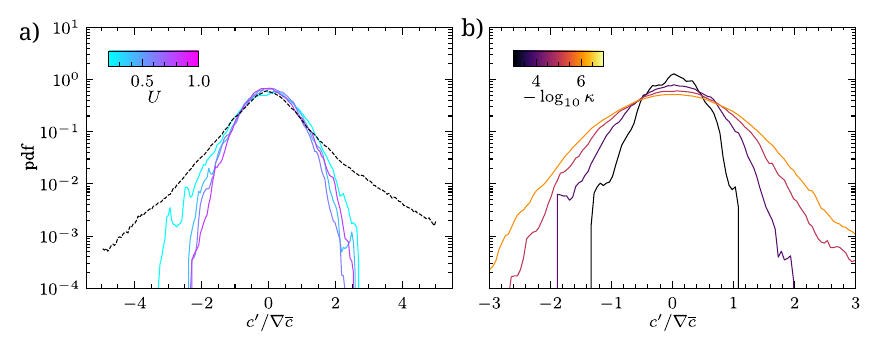} 
	\caption{Pdf of $c'$ as a function of $U$ (with $\kappa=10^{-5}$) and molecular diffusivity (with $U=0.4$). The black dotted line correspond to the case without large scale forcing, with the concentration normalized by $\sigma_c$. Pdf are ensemble averaged over 100 realisations.}
	\label{fig:pdf}
\end{figure}

Finally, we investigate the probability density function of scalar fluctuations for various flow amplitude (Fig.~\ref{fig:pdf}a) and molecular diffusivity (Fig.~\ref{fig:pdf}b). Due to the large-scale forcing, the pdf of fluctuation is stationary in time. The pdf shows a characteristic Gaussian shape, as previously observed in forced mixing cases~\citep{sinai1989limiting,Shraiman:Siggia}. The Gaussian shape is preserved for all flow persistence and molecular diffusivity studied.
Note also that the pdf of scalar fluctuations without large-scale forcing ($g=0$) have thicker power-law tails than the Gaussian distribution (Fig.~\ref{fig:pdf}a).
The Gaussianity of scalar fluctuations in the presence of a mean scalar gradient is an important feature because it allows one to accurately capture the distribution of fluctuations from the second moment only, that is, the scalar variance. 

\subsection{Scenario 2 : Fluctuations in a dispersing front}\label{sec:front}
\begin{figure}
    \centering
    \includegraphics[width=\linewidth]{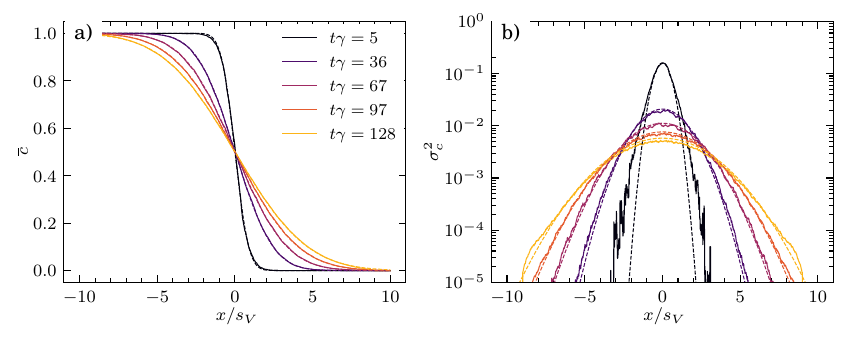}
    \caption{Direct numerical simulation  of the evolution the moments of a scalar front mixed by a chaotic velocity field (continuous line) in the scenario 2.  (a) Mean concentration and (b) Variance of the concentration (b). The velocity field is the sine flow with parameters $U=0.8$ and $s_v=T=1$. Moments are obtained by solving Eq.~\eqref{eq:ADE} with $\kappa=7.7 \cdot 10^{-7}$. The theoretical prediction using the closure Eq.~\eqref{eq:variance_front} is shown in dashed lines. The local mean and variance of concentration are obtained as $\bar{c}(x,t)=\langle c(x,y,t)\rangle_y$ and  $\sigma^2_c (x,t)= \langle c(x,y,t)^2\rangle_y - \bar{c}(x,t)^2$ respectively, where the average $\langle \cdot \rangle_y$ is taken on the periodic direction $y$ and on a ensemble of 100 independent realizations of the random phases $(\phi_i,\phi_j)$.}
    \label{fig:macrodisp}
\end{figure}

We now test the validity of Eq.~\eqref{eq:variance_closure} to describe the spatio-temporal evolution of the scalar variance in a unbounded mixing front, agitated by a chaotic microscale flow (the sine flow), and that spreads at large scales due to dispersion. An example of the instantaneous scalar field obtained in such scenario is shown in Fig.~\ref{fig:front}a.
Since $\bar{c}$ follows the macroscopic dispersion equation with $\overline{\boldsymbol{u}}=0$, its solution for an initial sharp step function is 
\begin{equation}
	\bar{c}=(1+\text{erf} (x/\sqrt{4Dt}) /2.
\end{equation}
As expected, with $D$ obtained from \eqref{eq:dispersion}, this solution accurately capture the numerical data averaged over 100 realisations of the random sine flow (Fig.~\ref{fig:macrodisp}a).
%
%
Using the closure model~\eqref{eq:variance_closure} and the independently estimated $s_i$ (Eq.~\eqref{eq:si_sine}), we can deduce the local value scalar variance. The mean scalar gradient is
\begin{equation}
	\frac{\partial \bar{c}}{\partial x}= \frac{1}{\sqrt{4\pi D t}} \exp(-x^2/(4Dt)),
\end{equation}
thus,
\begin{equation}
	\sigma^2_c(x,t)= \frac{ \exp(-x^2/(2Dt))}{ 2 \pi \gamma t}  \left| Ei( - 4\pi s_B'/s_i)\right|.
    \label{eq:variance_front}
\end{equation}
 Eq.~\eqref{eq:variance_front} accurately captures the evolution of scalar variance along the front at late time (Fig.~\ref{fig:macrodisp}b). At early times ($t\leq5$), the prediction slightly underestimates the spatial extent of the variance. This is probably due to the non-negligible transport of scalar variance at early time (the left hand side of Eq.~\eqref{eq:variance}), which redistribute fluctuations.
 
\begin{figure}
 	\centering
 	\includegraphics[width=0.75\linewidth]{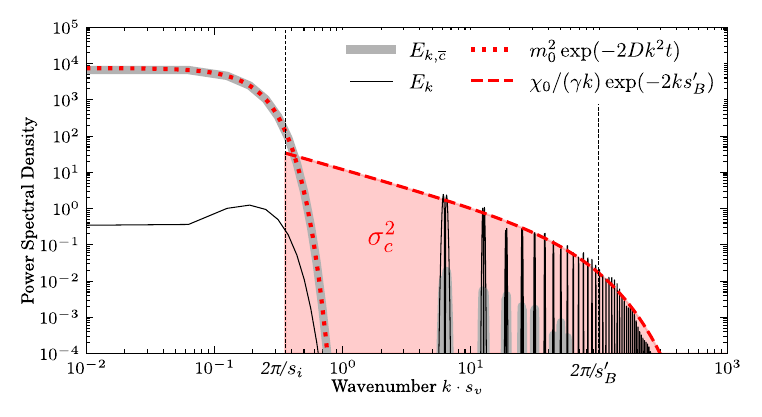}
 	\caption{Spectrum of the macroscale ($E_{k,\bar{c}}$) and microscale  ($E_{k}$) scalar fluctuations (thick and thin continuous lines respectively) in a freely dispersing front at time $T=1000$ in the 2D random sine flow for $U=0.4$ and $\kappa=10^{-5}$.  The spikes in the fluctuating part are caused by the periodicity of the sine flow. The theoretical dispersive spectrum at macroscale (Eq.~\eqref{eq:spectrum_dispersive} with $m_0$ the total mass of the scalar) and the microscale Kraichnan spectrum (Eq.~\eqref{eq:spectrum}, with $\chi_0/\gamma$ defined by Eq.~\eqref{eq:chi0_sine}) are shown in dotted and dashed lines respectively. In the wavenumber axis, we indicated the position of the injection scale $s_i=4\pi\sqrt{D\gamma^\prime/\gamma^2}$ and the modified Batchelor scale $s_B^\prime=\sqrt{\kappa/\gamma'}$. The shaded area determines the microscale scalar variance ($\sigma^2_c$)}
 	\label{fig:spectrum_sketch}
 \end{figure}

 The scalar PSD of macroscale ($E_{k,\bar{c}}$) and microscale ($E_{k}$) scalar fluctuations at late time ($T=1000$) in the mixing front are shown in Fig.~\ref{fig:spectrum_sketch}. For this simulation, we took a large domain length of $100 s_v$ to avoid boundary effects. The macroscale spectrum is well described by Fickian dispersion (Eq.~\eqref{eq:spectrum_dispersive}) while the microscale spectrum matches Kraichnan theory (Eq.~\eqref{eq:spectrum}) with the scalar variance injection rate driven by macroscale dispersion (Eq.~\eqref{eq:chi0_sine})). In turn, the injection scale $s_i$ is well identified in Fig.~\ref{fig:spectrum_sketch}, as an intermediate scale separating dispersive and stretching-enhanced mixing processes, and controlling the amount of scalar variance persisting in mixing fronts below macroscale. 
 
\section{Conclusion}
In this study, we have investigated the statistical behaviour of mixing fronts in single-scale chaotic flows--also termed smooth or laminar chaotic flows--with the objective of predicting the interactions between macroscopic scalar gradients driven by dispersion and microscale mixing controlled by fluid deformations. 

In time-uncorrelated chaotic flows, the statistics of concentration fluctuations are well described from the \citet{kraichnan1974convection} spectral theory of mixing below the characteristic velocity scale. In turn, above the velocity scale, the macroscopic evolution of concentration fields classically follow Fickian dispersion. 

We find that the characteristic scale $s_i$ at which macroscale dispersion and microscale mixing have the same dissipation rate can be used to match the two descriptions, and to describe the magnitude of microscale scalar fluctuations in large-scale dispersing fronts. The theoretical prediction~\eqref{eq:variance_closure} captures well the results of numerical simulations of mixing fronts in periodic sine flows for various Péclet numbers and flow persistences with no fitting parameters.  

The validity of the approach relies on three key properties of the flow. The first is the smoothness of the flow field, which ensures that stretching is a continuous, scale-independent process below the velocity-length scale and that dispersion is Fickian above the velocity scale. The second one is the scale separation induced by chaotic advection, which implies that the dissipation and production scales of the scalar variance are well separated, the Batchelor scale being well below the characteristic velocity length scale. The third is the rapid time-decorrelation of flow, which leads to the Kraichnan microscale spectrum. However, we show that a finite amount of flow persistence ($Ku\sim 1$) does not strongly affect the quality of the prediction for $Pe>200$. 
The closure \eqref{eq:variance_closure} is thus expected to apply in any smooth single-scale chaotic flow of moderate persistence. This is the case for laminar flows, for instance in steady flow through 3D porous media, where the velocity fluctuations induce chaotic advection in the 2D plane transverse to mean flow direction~\citep{Lester2013}. In such flows, fluid deformations occur on a time scale that is related to the velocity length scale and magnitude ($T \sim s_v/U$), such that Eq.~\eqref{eq:stretching_def} simplifies to $\gamma\sim U/s_v$. Thus, flow persistence is of order $Ku\sim 1$.

Our study shows that microscale scalar fluctuations persist at late time in chaotically mixed fronts. However, even at large Péclet numbers, they remain limited compared to macroscale scalar gradients. For example, at $Pe=10^3$, scalar fluctuations are expected to be less than 10\% of the macroscopic gradient (Fig.~\ref{fig:variance}a).  At $Pe=10^8$, the fluctuations hardly reach 60\% of the macroscopic gradient. This observation is reminiscent of the efficient nature of chaotic flows to mix, even at high Péclet numbers.
Because it describes the amount of mixing at microscale in a macroscopically dispersing front, the proposed closure opens a new avenue to upscale non-conservative transport processes, such as reactive processes. Extension to bimolecular reactive transport system can be obtained by focusing on the concentration statistics of two reactive species, with fluctuation product $\overline{c_A' c_B'}$ rather than a single one $\overline{c'c'}$ as done here~\citep{anmala2013dynamics}. 

Additional work is required to explore how the specific value of $s_i$ changes for other single-scale flows than the sine flow, or for multiscale laminar flows. 
Finally, it would be interesting to compare the value of $s_i$ to the concept of \textit{coarsening} scale, first proposed by \citet{villermaux2003mixing} for turbulent flows. The coarsening scale was defined as an intermediate length scale, larger than $s_B$, below which scalar fluctuations are quickly erased by turbulent agitation. As shown in Fig.~\ref{fig:spectrum_sketch}, $s_i$ indeed delimits an  abrupt drop in the scalar spectrum and may thus be considered as a natural coarsening scale, separating dispersion- from stretching-induced scalar features. Since the coarsening scale was initially introduced in the context of multiscale turbulent flows \citep{rotily2025momentum}, the applicability of this concept to rough chaotic flows remains to be established.
%

%


\end{document}